\documentclass[a4paper,12pt]{ifsa2001}
\usepackage{graphicx}
\usepackage{times}      % to use Postscript font (PSNFSS system)

\input{ifsa2001.mac}

\title{
 {\rm\footnotesize Proc. of the 2nd Int. Conference on Inertial Fusion Science and Applications,
                    IFSA2001, Kyoto, Japan, Sept. 9-14, 2001.
        Edited by K.~A.~Tanaka, D.~D.~Meyerhofer, J.~Meyer-ter-Vehn (Paris: Elsevier, 2002) pp. 106-109 }
 \vskip -2mm
 \rule{160mm}{0.15mm}
 \vskip  3mm
 \bf Molecular Dynamics Simulation on Stability of Converging Shocks}
\author{\vspace{3mm} {\rm V.~Zhakhovskii, K.~Nishihara, and M.~Abe}\\
{\it Institute of Laser Engineering, Osaka University,}\\
{\it 2-1 Yamada-oka Suita,Osaka, 565, Japan}
 }

\begin{document}
 \thispagestyle{empty}
 \maketitle
 \thispagestyle{empty}
\begin{abstract}

Molecular Dynamic (MD) approach is applied to study the converging cylindrical shock waves in a dense
Lennard-Jones (LJ) fluid. MD method is based on tracking of the atom motions and hence it has an fundamental
advantages over hydrodynamic methods which assumes shocks as a structureless discontinuity and requires an
equation of state. Due to the small thickness of shock fronts in liquid the two million particles is enough to
simulate propagation of a cylindrical shocks in close detail.

We investigate stability of converging shocks with different perturbation modes and its mixture. It was shown
that in a case of relatively large initial ripples the Mach stems are formed. Supersonic jets generated by
interaction of reflected shocks in downstream flow are observed. We also study the Richtmyer-Meshkov (RM)
instability of an interface between two Lennard-Jones liquids of different mass densities. Surprisingly, mode 3
ripples grow very slow in comparison with higher mode numbers and growth rate of a higher mode decay slower.

\end{abstract}

% ----------------------------------------------
 \vspace{-1mm}
 \section{Introduction}
 \vspace{-0mm}

Molecular Dynamic simulations involving the converging shock stability and the Richtmyer-Meshkov fluid
instability at atomic scale are presented for the first time. Understanding of these instabilities is very
important in inertial fusion science. So far, the capacity of the fluid numerical methods to compute fluid flow
at small lengths calls in question, especially in case of converging shocks \cite{Yang}.

Last 40 years MD method was successfully applied to many problems in different fields such as statistical and
chemical physics. With increasing of the computer power, simulations of many millions of atoms in 3D space are
now feasible. We believe that nowadays is time to apply MD method to fluid hydrodynamic problems.

MD approach is based on tracking of the atom motions and hence it can provide us whole information of a system
behavior including phase transition in the matter, diffusion, thermoconductivity and other nonequilibrium
processes far from thermodynamic steady state. Due to direct computation of the atom movements and the
atom-atom interactions MD method has an fundamental advantages over numerical hydrodynamic methods which
assumes shocks as a structureless discontinuity and requires an equation of state. Moreover, the different
hydrodynamic codes show significant deviations in the simulation flow and a higher mesh resolution can not
improve the situation \cite{Kamm}, but the various MD programs give the same simulation results.

Fluid simulations of the Richtmyer-Meshkov instability by using numerical hydrodynamic methods have so far
failed to provide quantitative agreement with experiments due to grid induced numerical instability. For this
reason, it is necessary to apply an alternate method, namely MD approach as a very promising method of
hydrodynamic simulation. It is the purpose of this work to carry out such numerical experiments. Simulations
were performed by own full vectorized MD code -- Molecular Dynamics Solver ( 11 seconds per one time step for
$2^{21}$ LJ atoms, and 92 ns per a pair of atoms on the NEC SX-5 supercomputer ).

\begin{figure}[ht]           %  Figure 1
  \centering
    \includegraphics [width=160mm]{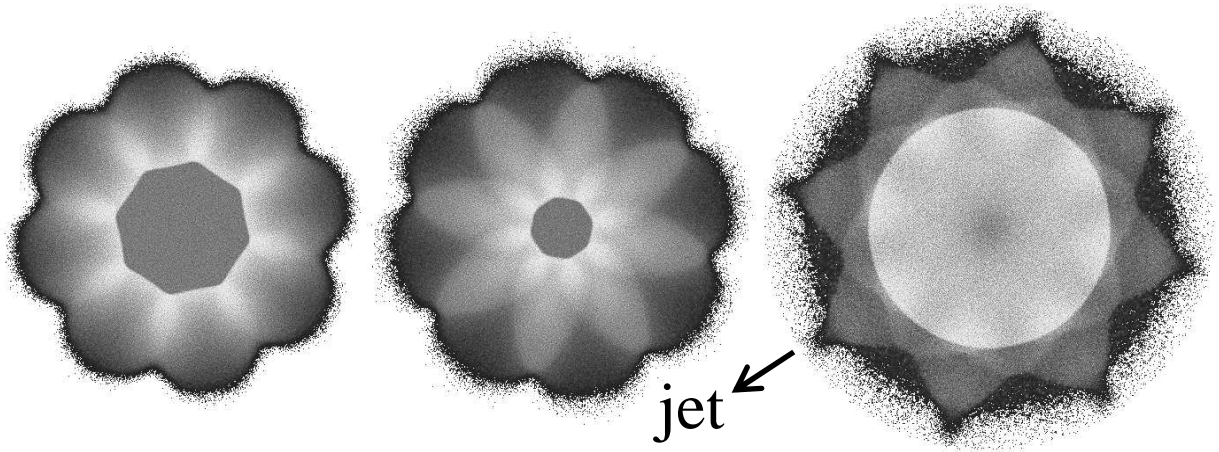}
    \caption{\hspace{2mm} MD simulation of perturbed shock front at $m=8$ and
     $v_p=1.25$. Brightness corresponds to atom density, each pixel is occupied by
     20 atoms roughly. ({\it left},$t=38.6$): First generation of Mach stems appears
     at very early stage of converging SW. ({\it center},$t=50.5$): Phase inversion
     of the perturbations occurs.({\it right},$t=96.5$): Reflected shock has almost
     unperturbed symmetrical front and 8 supersonic jets run away from the target
     surface. Animation is located on
     {\sl www.ile.osaka-u.ac.jp/research/TSI/Vasilii/}~~, file nd-pb08.gif, size 5.1 MB. }
\end{figure}

\vspace{-1mm}
\section{Simulation technique and results}
\vspace{-0mm}

In our MD simulation the atoms of a liquid interact via modified Lennard-Jones short-range pair potential with
cut-off and smoothing at $r_{cut}=2.5\sigma$ \cite{basil99}:
\begin{eqnarray}
\phi_{LJ}(r) &=&
  4\epsilon\left[\left(\sigma/r\right)^{12}-\left(\sigma/r\right)^6 \right]-
  a_{2}(r^{2}-r_{0}^{2})^{2}-a_{3}(r^{2}-r_{0}^{2})^{3} ,
   \quad \hbox{\rm if} \quad 0 < r \leq r_{cut} , \label{LJsmooth} \nonumber
\end{eqnarray}
where $r_{0}=2^{1/6}\sigma$, and $\sigma$,$\epsilon$ are the usual LJ parameter, and $a_{2}=-3.5289\times
10^{-3}$,$a_{3}=5.75868\times 10^{-4}$. We use these parameters and the atomic mass $m_a/48$ as reduced
molecular dynamic units. Here and after all quantities are in MD units (mdu). For argon atoms $\sigma\!=
3.405\,$\.{A}, $\epsilon /k_{B}=119.8$ K,  time mdu$=\sigma \sqrt{m_a/48\epsilon}=3.113\times 10^{-13}$s,
velocity mdu $=1094$ m/s, and the unit of pressure is $0.0419\,{\rm GPa}$.

\begin{figure}[h]             % Figure 2
  \centering
    \includegraphics [width=160mm]{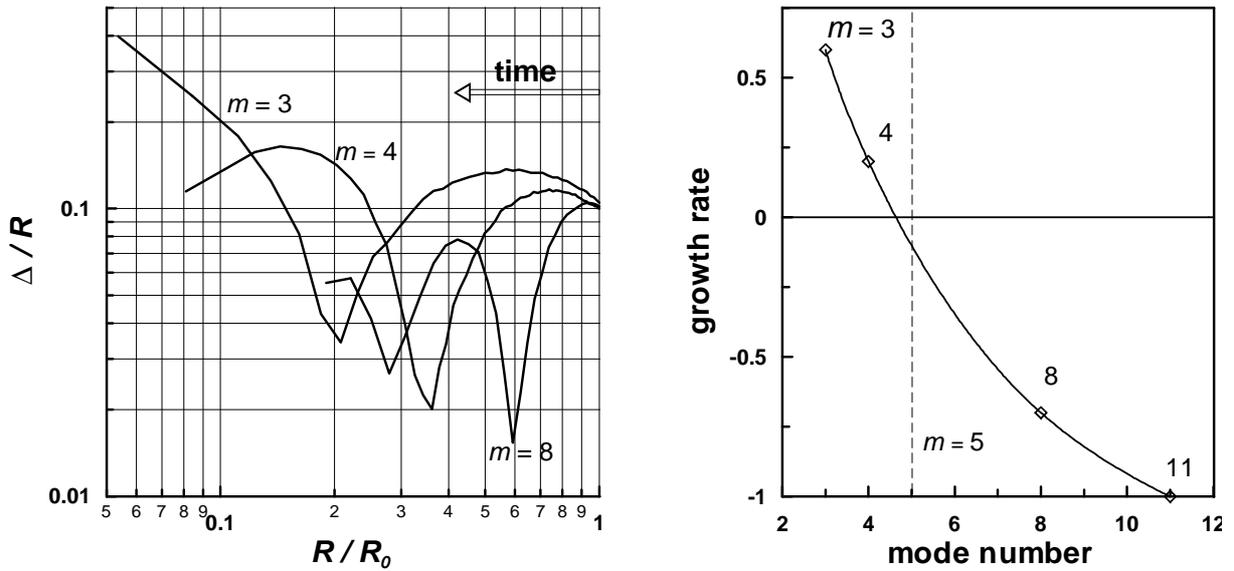}
    \caption{\hspace{2mm} ({\it left}): Normalized perturbation amplitudes for
                  different mode numbers, as a function of the shock position.
                  ({\it right}): Growth rate of perturbations vs. mode number.}
\end{figure}

At the beginning atoms are placed into rectangular MD cell inside cylindrical domain surrounded by a piston as
a cylinder wall. The piston is simulated by an external potential $\sim [r-R(\phi,t)]^2$ and its position
depends on angles in case of a perturbed boundary as well as on time to generate shock waves. The total number
of atoms in simulation is $2097148$. The computational MD cell has dimensions $L_x\times L_y\times L_z$, where
$L_x = L_y=549.7$ and the thickness of MD cell is $L_z=21.8$, and periodical boundary conditions are imposed on
the system along the cylinder axis (z-axis).

At the preparation stage the cylinder radius is fixed $R_0=197.25$ and all atoms have the same atomic weight
$m_a=48$. The Langevin thermostat is applied to prepare initial cold LJ liquid at the thermodynamic equilibrium
with given temperature $T=0.72$ and number density $n=0.79$. In case of a perturbed cylinder boundary the
initial position of the piston is defined as $R(\phi) = R_0[1+ \delta \sin(m\phi)]$ where $m$ is the mode
number, and $\delta=0.05$ is the relative perturbation, $\Delta=2\delta R_0$ is the initial perturbation
amplitude.

In the end of the preparation stage an equilibrium state with uniform mass density is reached. At the beginning
of the simulation stage in order to simulate two different materials the atoms outside of the interface change
instantly its mass to heavy one $m_b=16m_a$ and its velocities to $v_b=v_a/4$ to hold uniform temperature and
pressure. Interaction between light-heavy atoms and heavy-heavy atoms is described by the same LJ potential (1)
without any changing of the parameters. At $t=0$ the piston starts to move with the constant velocity $v_p$ and
at $t=16$ the piston is removed from the system. We estimate the sound velocity in light material as $c=0.7$
and measured shock Mach number is $M=3.3$ for piston velocity $v_p=1$.

\begin{figure}[h]           % Figure 3
  \centering
    \includegraphics [width=160mm]{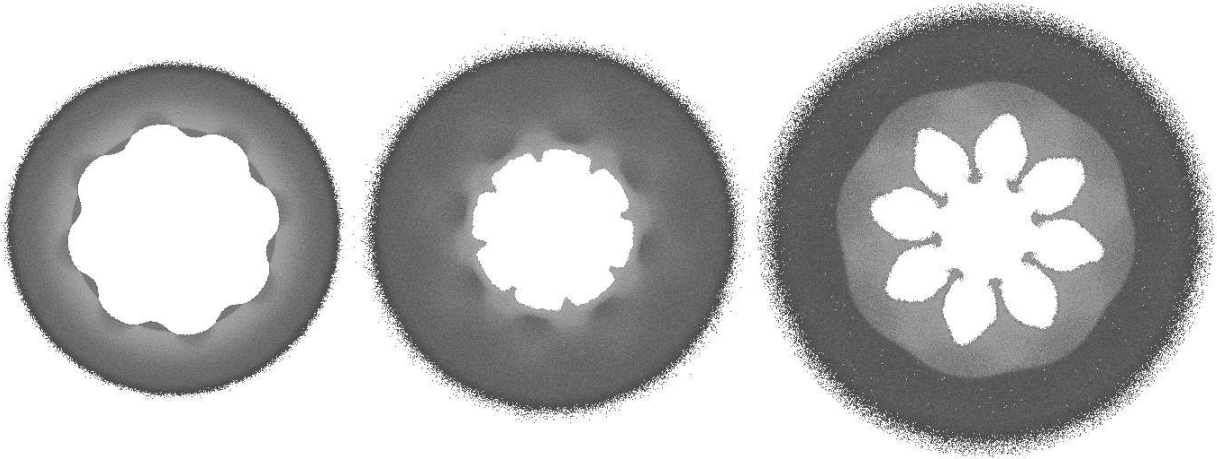}
    \caption{\hspace{2mm} MD simulation of RM instability for $m=8$ and $v_p=1$.
           Brightness corresponds to atom density, only heavy material is shown.
           ({\it left},$t=68.3$): Shock front hits material interface. Reflected wave is
           rarefaction wave.
          ({\it center},$t=139.5$): Spikes and bubbles begin to grow after hitting of
           the interface by SW reflected from origin (see Fig.5, second triangle)
          ({\it right},$t=267.2$): Mixing zone mounts almost the asymptotic stage.
             Animation is located on  {\sl www.ile.osaka-u.ac.jp/research/TSI/Vasilii/}~~,
             nd-rm08.gif, size 8.2 MB.}
\end{figure}
\begin{figure}[h]             % Figure 4
  \centering
    \includegraphics [width=160mm]{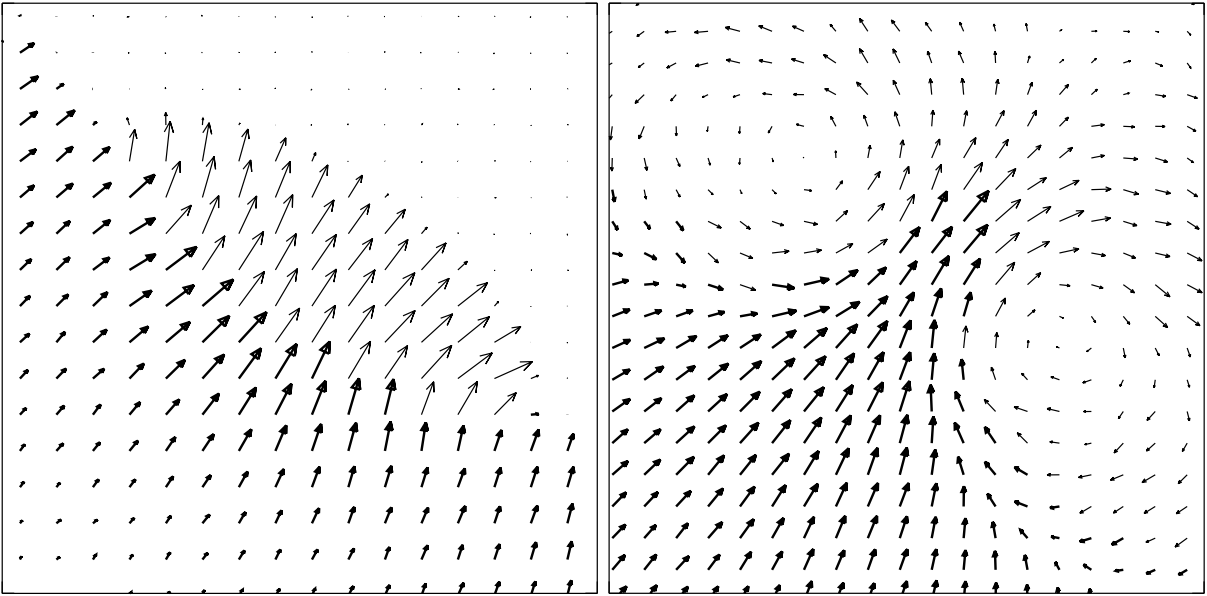}
    \caption{\hspace{2mm} Flow velocity map at heavy-light interface in case of the
            RM instability for $m=8$. Dark arrows correspond to heavy material, light arrows -
            light material. Corresponding density maps are shown in Fig.3.
            ({\it left},$t=68.3$): SW-interface interaction. Each arrow is an averaging
            of 320 atom velocities roughly. Frame size is 70 mdu.
            ({\it right},$t=139.5$):  Flow velocity map around spike. Each arrow represents
            the mean velocity about 100 atoms. Frame size is 40 mdu.}
\end{figure}
\begin{figure}[h]             % Figure 5
  \centering
    \includegraphics [width=160mm]{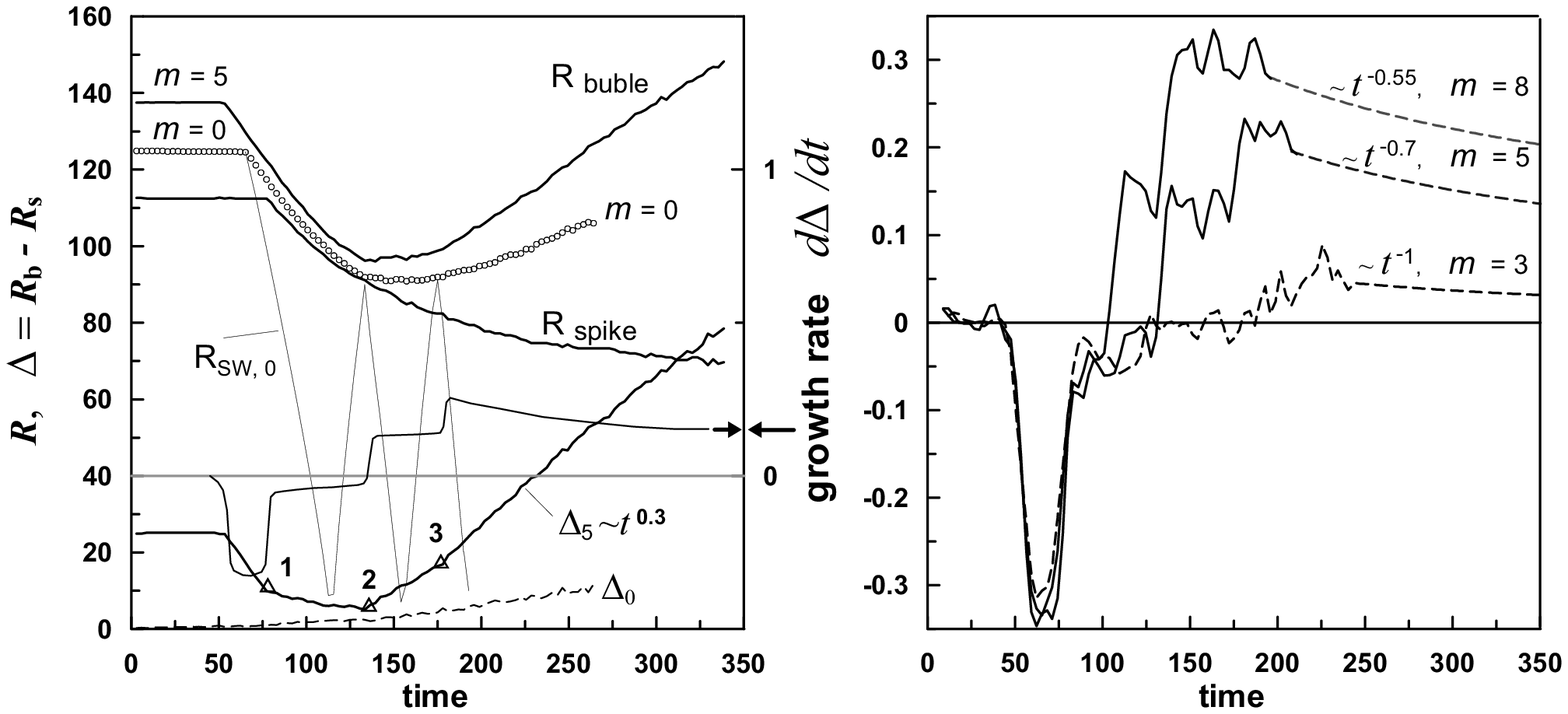}
    \caption{\hspace{2mm} Time history of the RM instability. $R_b$ and $R_s$ are
    maximum and minimum of interface location. At the late time, after 2nd
    triangle, they correspond to the radial position of the bubbles and spikes.
    ({\it left}): Circles denote interface position in symmetrical case without any
    ripples. Trajectory of the cylindrical shock front $R_{SW,0}$ is shown because of
    perturbed shock has a close location on average.
    Triangles correspond to key points of SW-interface interaction (see text).
   ({\it right}): Growth rates of perturbations vs. time for different mode number.}
\end{figure}

Typical snapshots of MD system in the case of the perturbed converging shock are shown in Figure 1. It is found
that for enough large initial perturbation the curved shock front generates Mach stems at very early stage of
converging. Then the first generation of Mach stems entails the secondary Mach stems and so on. In fact
polygons converge to the center instead of smooth curved shock front. There are $m$-side polygons for mode $m$
or $2m$-sides at phase inversion stage. Also the Mach stems generate intricate structures of reflected shocks
in downstream flow which result in supersonic jets running away from the target boundary. We suppose these jets
can amplify instability of ablation front in the real ICF experiments. Figure 2 shows amplitudes of the
measured shock front perturbation as a function of a mean shock radius and the perturbation growth rate for
different mode numbers. The simulation results show that the perturbations grow if the mode number $m<5$
\cite{Gard82}.

Figures 3 and 4 represent simulation results of the RM instability where SW propagates from heavy to light
liquids. The typical flow velocity fields are presented in Fig.4. When the shock front hits the interface, the
transmitted shock and rarefaction wave are produced and go away from the material interface (Figs. 3,4:{\it
left}). It is clearly seen that the vorticities are allocated around the material interface and they cause a
spike to be formed (Fig.4:{\it right}). Fig.3:{\it right} shows the RM instability at late simulation time when
the total width of the mixing zone goes into the asymptotic stage of the growth.

The time history of the RM instability is shown in Figure 5. We measure the thickness of the mixing zone
$\Delta$ as a distance between the maximum and minimum of interface positions along the radius. For comparison
with the symmetrical case we draw mean position (circles) of the cylindrical interface $m=0$ as a function of
time. It should be mentioned that small atomic perturbations with $m\gg 1$ even in a pure symmetrical case are
unavoidable due to atomistic fluctuations and they begin grow as well as artificial ripples (see $\Delta_0$ in
Fig.5 {\it left}).

There are three key time points in the RM instability evolution. The first point (1st triangle in the left
graph of Fig.5) corresponds to a contact of the shock front with a bottom of the ripple. After that the maximum
and minimum are closing in time. The phase inversion occurs at the minimum of $\Delta$, between 1st and 2nd
triangles. For the higher mode number the inversion takes place at earlier time than that for the lower mode
number. The shock reflected at the origin reaches the interface at the second key time point and it accelerates
the growth of the interface perturbations. In the case of mode 5 the coincidence of the phase inversion with
the second point occurs accidentally, the phase inversion of the interface for $m=3$ takes place after the
secondary reflected shock pushes the interface outward at 3rd key time point, $t\approx 180$.

Figure 5:{\it right} indicates nonlinear evolution of the growth rates for the different modes. Due to
numerical differentiation of the thickness of the mixing zone $\Delta=\Delta(t)$, the growth rate $d\Delta/dt$
is noisy and we are forced to estimate asymptotical behavior of the growth rate from the differentiation of
$\Delta(t)$ asymptotic form. In the case of mode 8 the perturbations grow as $\Delta_8 \sim t^{0.45}$, and
$\Delta_5 \sim t^{0.3}$, $\Delta_3 \sim \ln(t)$. Hence the growth rates decay $d\Delta_8/dt \sim t^{-0.55}$,
and $d\Delta_5/dt \sim t^{-0.7}$, and $d\Delta_3/dt \sim t^{-1}$ correspondingly.

\vspace{-4mm}
\section{Conclusion}
\vspace{-1mm}

Molecular dynamics simulation of cylindrical shocks have been performed for the first time. We have
demonstrated that MD method is able to catch hydrodynamic instabilities in close details on atomic scale.

It is shown that for a converging cylindrical shock $M \approx 3.3$,  the compression of Argon liquid is 2.2 at
the collapse point, the temperature is around 12000 K, pressure is $\approx 50$GPa, converging time $ \approx
20$ps. The growth rate of a ripples strongly depends on the perturbation modes and slowly decreases in time.
Perturbations grow faster for higher mode in the case of the RM instability. Mode 3 ripples grow very slow in
comparison with higher mode numbers due to the phase inversion of the interface occurs at very late time.

\end{document}